\documentclass[12pt]{article}
\usepackage{epsf}
\usepackage{amsmath}
\usepackage{graphics}

\renewcommand{\thefootnote}{\#\arabic{footnote}}

\begin{document}

\setcounter{footnote}{0}
\begin{titlepage}

\begin{center}

\hfill astro-ph/0511821\\
\hfill November 2005\\

\vskip .5in

{\Large \bf
Dark Energy Evolution and the Curvature \\
of the Universe from  Recent Observations
}

\vskip .45in

{\large
Kazuhide Ichikawa and Tomo Takahashi
}

\vskip .45in

{\em
Institute for Cosmic Ray Research, \\
University of Tokyo, Kashiwa 277-8582, Japan
}

\end{center}

\vskip .4in

\begin{abstract}

We discuss the constraints on the time-varying equation of state for
dark energy and the curvature of the universe using observations of
type Ia supernovae from Riess et al. and the most recent Supernova
Legacy Survey (SNLS), the baryon acoustic oscillation peak detected in the
SDSS luminous red galaxy survey and cosmic microwave background. Due
to the degeneracy among the parameters which describe the time
dependence of the equation of state and the curvature of the universe, the
constraints on them can be weakened when we try to constrain them
simultaneously, in particular when we use a single observational data.
However, we show that we can obtain relatively severe constraints when
we use all data sets from observations above even if we
consider the time-varying equation of state and do not assume a flat
universe.  We also found that the combined data set favors a flat
universe even if we consider the time variation of dark energy
equation of state.

\end{abstract}
\end{titlepage}

\renewcommand{\thepage}{\arabic{page}}
\setcounter{page}{1}
\renewcommand{\thefootnote}{\#\arabic{footnote}}

\section{Introduction}

Almost all current cosmological observations indicate that the present
universe is accelerating. It can be explained by assuming that
the universe is dominated by dark energy today. Although many candidates for
dark energy have been proposed so far, we still do not know the nature
yet.  Many studies have been devoted to investigate dark energy
assuming or constructing a specific model and then study its
consequences on cosmological observations such as cosmic microwave
background (CMB), large scale structure (LSS), type Ia supernovae
(SNeIa) and so on.  On the other hand, many efforts have also been
made to study dark energy in phenomenological way, i.e., as model
independent as possible.  In such approaches, dark energy can be
parameterized with its equation of state and constraints on it can be
obtained using cosmological observations.  Assuming that the equation
of state for dark energy $w_X$ is constant in time, current
observations give the constraint as $w_X \sim -1$
\cite{Spergel:2003cb,Tegmark:2003ud,Tonry:2003zg,Riess:2004nr,MacTavish:2005yk,Astier:2005qq}.
Although one of the most famous models for dark energy is the
cosmological constant whose equation of state is $w_X = -1$, most
models proposed so far have time-varying equation of state.  Thus,
when we study dark energy, we should accommodate such time dependence
in some way.

Many recent works on dark energy investigate the time dependence of
the dark energy equation of state using simple parameterizations such
as $w_X = w_0 + w_1 (1-a)$ with $a$ being the scale factor of the
universe \cite{Chevallier:2000qy,Linder:2002et}.  Parameterizing the
dark energy equation of state in simple ways, the constraints on 
the time evolution of $w_X$ have been considered (for recent works on this issue, for
example, see Ref.~\cite{w_const}).  It should be mentioned that, when
one studies the equation of state for dark energy, it is usually
assumed that the universe is flat.  It should be also noted that dark
energy is usually assumed to be the cosmological constant when one
derives the constraint on the curvature of the universe.  However, it
has been discussed that, assuming a non-flat universe, the constraints
on $w_X$ and the curvature of the universe can be relaxed to some
extent even with the time independent $w_X$ from the CMB data alone
\cite{Crooks:2003pa}.  Also, even if we assume a flat universe, there
are degeneracies among the parameters which describe the evolution of
dark energy, i.e., the time dependence of $w_X$ when we consider the
constraints on dark energy \cite{Maor:2000jy,Maor:2001ku}.
Furthermore it has been also discussed that if we remove the prior of
a flat universe, the degeneracies becomes much worse
\cite{Wang:2004py}.  Since it is very important to study the time
dependence of $w_X$ to differentiate the models of dark energy and
also the curvature of the universe to test the inflationary paradigm,
we should investigate how the prior on the curvature of the universe
affects the determination of the time-varying equation of state for
dark energy and vice versa.  Some works along this line have been done
using a specific one-parameter parameterization for the time-varying
$w_X$ \cite{Gong:2005de}.

In this paper, we study this issue, namely the determination of the
evolution of dark energy and the curvature of the universe, in some
detail using widely used parameterization of the equation of state for
dark energy. We consider the constraints from observations of SNeIa
reported in Refs.~\cite{Riess:2004nr,Astier:2005qq}, the baryon acoustic
oscillation detected in the SDSS luminous red galaxy survey
\cite{Eisenstein:2005su} and recent CMB observations including WMAP
\cite{Spergel:2003cb}.  In the next section, we summarize the analysis
method we adopt in this paper. Then we discuss the constraints from
above mentioned observations on the curvature of the universe with
time-varying equation of state for dark energy followed by the
analysis on $w_X$ without assuming a flat universe.  The final section
is devoted to the summary of the paper.

\section{Method}

In this section, we briefly summarize the method for constraining the
parameters which describe the dark energy evolutions and other
cosmological parameters.  To study the evolution of dark energy, we
use the following parameterization for the time-varying equation of
state \cite{Chevallier:2000qy,Linder:2002et}
\begin{equation}
w_X (z) = w_0 + \frac{z}{1+z} w_1 = w_0 + (1-a) w_1,
\label{eq:eos}
\end{equation}
where $z$ is the redshift.  In this parameterization, the equation of
state at the present time is $w_X (z=0) = w_0$ and for the early time
it becomes $w_X(z=\infty) = w_0+w_1$.  Since we are interested in the
late-time acceleration of the universe due to dark energy, we consider
the case where the dark energy dominates the universe only at late
time.  Thus, in this paper, we assume
\begin{equation}
w_0+w_1 < 0, 
\label{eq:w0w1}
\end{equation}
in order not to include the possibilities of early-time dark energy
domination.  With this parameterization, the energy density of dark
energy can be written as
\begin{equation}
\rho_X(z) = \rho_{X0}  (1+z)^{3(1+w_0+w_1)}  \exp \left( \frac{-3 w_1 z}{1+z} \right),
\end{equation}
where $\rho_{X0}$ is the energy density of dark energy at present
time.  The Hubble parameter is given by
\begin{equation}
H^2 (z) = H_0^2 \left[  \Omega_r (1+z)^4
+\Omega_m (1+z)^3 + \Omega_k (1+z)^2 
+ \Omega_X  (1+z)^{3(1+w_0+w_1)}  \exp \left(  \frac{-3 w_1 z}{1+z}  \right) \right],
\end{equation}
where $H_0$ is the Hubble parameter at the present epoch, $\Omega_i$
is the energy density of a component $i$ normalized by the critical
energy density and the subscripts $r, m,k$ and $X$ represent
radiation, matter, the curvature of the universe and dark energy,
respectively.  To consider the constraints on dark energy and other
cosmological parameters, we use the data from SNeIa, the baryon
acoustic oscillation peak and the CMB.

As for SNeIa data, we use the gold data set given in
Ref.~\cite{Riess:2004nr} and the first year data of the Supernova
Legacy Survey (SNLS) released recently \cite{Astier:2005qq}.
Constraints from SNeIa can be obtained by fitting the distance modulus
which is defined as
\begin{equation}
M -m = 5\log \left( \frac{d_L}{\rm Mpc}  \right)+ 25. 
\end{equation}
Here $d_L$ is the luminosity distance which is written as
\begin{equation}
d_L = \frac{1+z}{\sqrt{|\Omega_k|} }
\mathcal{S}  \left( \sqrt{|\Omega_k|} \int_0^z \frac{dz'}{H(z') /H_0} \right),
\end{equation}
where $\mathcal{S}$ is defined as $\mathcal{S}(x) = \sin (x)$ for a
closed universe, $\mathcal{S}(x) = \sinh (x)$ for an open universe and
$\mathcal{S}(x) = x$ with the factor $\sqrt{|\Omega_k|}$ being removed
for a flat universe.

We also use the baryon acoustic oscillation peak detected in the SDSS
luminous red galaxy survey \cite{Eisenstein:2005su}. To obtain the
constraint, we make use of the parameter $A$ which is defined as
\begin{equation}
A = \frac{\sqrt{\Omega_m}}{(H(z_1)/H_0)^{1/3}} \left[  
\frac{1}{z_1 \sqrt{|\Omega_k|} }
\mathcal{S}  \left( \sqrt{|\Omega_k|} \int_0^{z_1} \frac{dz'}{H(z')/H_0} \right)
\right]^{2/3},
\end{equation} 
where $z_1=0.35$ and $A$ is measured to be $A=0.469\pm 0.017$
\cite{Eisenstein:2005su}.

For the CMB data, we only use the shift parameter $R$ which determines
the whole shift of the CMB angular power spectrum
\cite{Bond:1997wr}. $R$ is given by
\begin{equation}
R  = \frac{\sqrt{\Omega_m}}{\sqrt{|\Omega_k|} }
\mathcal{S}  \left( \sqrt{|\Omega_k|} \int_0^{z_2} \frac{dz'}{H(z')/H_0} \right),
\end{equation}
where $z_2=1089$.  It has been discussed that using the CMB shift
parameter is a robust way to include the constraints from observations
of CMB \cite{Wang:2003gz}.  From the recent observations of CMB
including WMAP, CBI and ACBAR, the shift parameter is constrained to
be $R = 1.716 \pm 0.062$ \cite{Spergel:2003cb,Wang:2003gz}

In this paper, we only consider the effect of the modification of the
background evolution by the change in the cosmological parameters. In
fact, the properties of dark energy can also modify the evolutions of
cosmic density fluctuation.  When we consider the effects of dark
energy perturbation, we also have to specify the speed of sound of
dark energy component.  However, such modification can arise at low
multipole region of the CMB power spectrum where the errors due to
cosmic variance are large. Thus the constraints from CMB on dark
energy mostly come from the position of acoustic peaks which can be
described by the shift parameter.  Hence we do not consider the
perturbation of dark energy in this paper\footnote{
However, the speed of sound can be useful to differentiate the models
of dark energy.  There are some works which discuss the constraint on
the speed of sound.  For interested reader, we refer
Refs.~\cite{Bean:2003fb,Weller:2003hw,Hannestad:2005ak}. }
.

\section{Constraints on the $\Omega_m$ vs. $\Omega_X$ plane}

\begin{figure}[t]
    \centerline{\scalebox{1.2}{\includegraphics{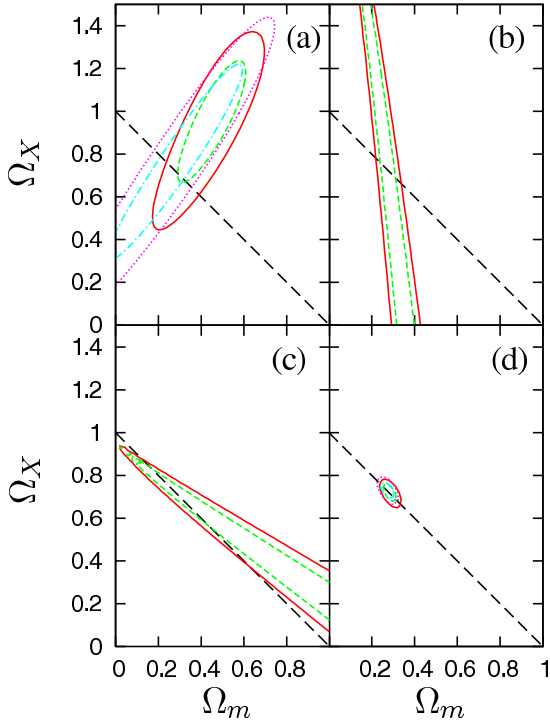}}}
	\vspace{0.5cm} \caption{Constraints on $\Omega_m$ and
	$\Omega_X$ from SNeIa (a), the baryon acoustic oscillation peak
	(b), the CMB shift parameter (c) and all data combined (d).
	For the constraint from SNeIa data and all data combined, we
	show the contours obtained from the gold set of
	Ref.~\cite{Riess:2004nr} and SNLS \cite{Astier:2005qq}
	separately. Contours of 1$\sigma$ (red solid line) and
	2$\sigma$ (green dashed line) are shown (for SNLS data,
	1$\sigma$ and 2$\sigma$ contours are shown in blue dash-dotted
	line and purple dotted line, respectively).  We assumed the
	cosmological constant as dark energy and a flat universe
	here.}
\label{fig:Om_Ox_wm1}
\end{figure}

Now we discuss the implication of the dark energy evolution on the
determination of the curvature of the universe.  For this purpose, we
derive the constraints from the observations on the $\Omega_m$
vs. $\Omega_X$ plane.  First we consider the case with the constant
equation of state for dark energy. In Fig.~\ref{fig:Om_Ox_wm1}, we
show the contours of 1$\sigma$ and 2$\sigma$ constraints from
observations of SNeIa (a), the baryon acoustic oscillation peak (b)
and the CMB shift parameter (c).  We also show the constraint from the
combination of all data sets (d).  In Fig.~\ref{fig:Om_Ox_wm1}, we
assume the equation of state as $w_X=-1$.  For the constraints from
SNeIa, we used the gold data set from Ref.~\cite{Riess:2004nr} and the
data from SNLS \cite{Astier:2005qq} separately.  As we can see,
although each constraint from a single observational data has the
degeneracy in the $\Omega_m$ vs. $\Omega_X$ plane, when all the data
sets are combined, we can obtain a severe constraint as $\Omega_m \sim
0.3$ with $\Omega_k \sim 0$, which is a well-known result.

\begin{figure}[t]
    \centerline{\scalebox{1.2}{\includegraphics{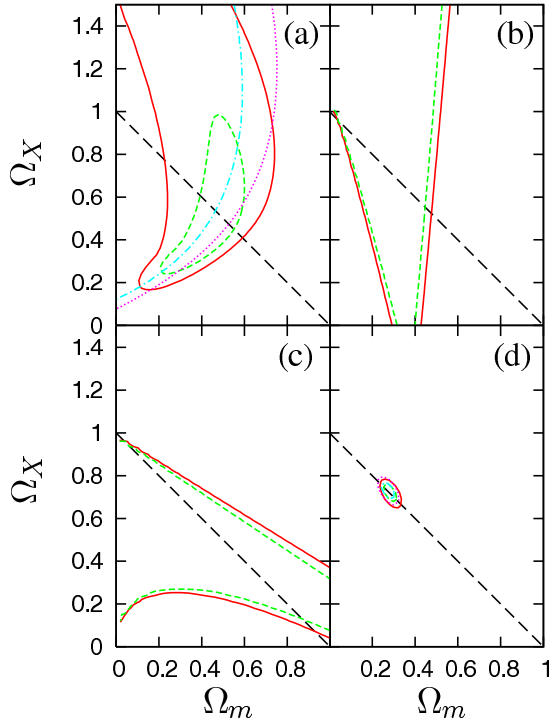}}}
	\caption{The same as Fig.~\ref{fig:Om_Ox_wm1} except that we
	marginalized over the value of $w_X$. Here we considered the
	constant $w_X$. }
\label{fig:Om_Ox_wmarg}
\end{figure}

\begin{figure}[t]
    \centerline{\scalebox{1.2}{\includegraphics{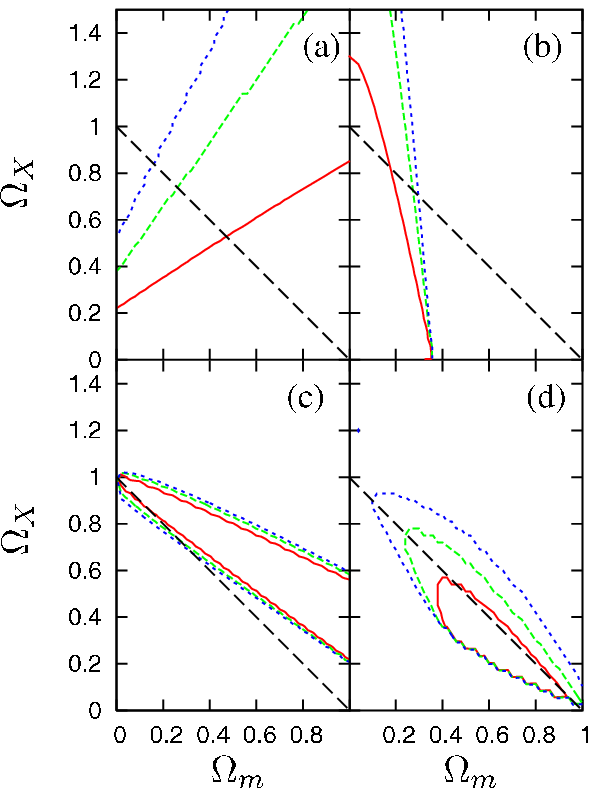}}}
     \vspace{0.5cm}
	\caption{Contours of constant $w_X$ which gives minimum
	$\chi^2$ when we marginalize over $w_X$ for the constraints
	from SNeIa (a), the baryon acoustic oscillation peak (b), the CMB
	shift parameter (c) and all data combined (d).  Contours of
	$w_X=-2$ (red solid line), $-1$ (green dashed line) and $-0.8$
	(blue dotted line) are shown except for the case where all
	three data are combined (panel (d)). For the panel (d),
	contours of $w_X=-1.1$ (red solid line), $-1$ (green dashed
	line) and $-0.9$ (blue dotted line) are shown.  For SNeIa, the
	data from SNLS are used here.}
\label{fig:Om_Ox_wmarg_w0min}
\end{figure}

Next we consider the case where the equation of state for dark energy
is allowed to vary, but still we keep $w_X$ constant in time.  In
Fig.~\ref{fig:Om_Ox_wmarg}, we show the contours of 1$\sigma$ and
2$\sigma$ constraints after marginalizing over the values of $w_X$.
Here we assumed the prior on $w_X$ as $ -5 \le w_X \le 0$.  As seen
from the figure, the allowed regions become larger compared to those
for the case with $w_X=-1$ when we use a single data set
alone. However, by combining all data, we can obtain almost the same
constraint as the case with $w_X=-1$ even though we marginalize over
$w_X$.  This can be understood by noting that the values of $w_X$
which give the minimum $\chi^2$ for fixed $(\Omega_m, \Omega_X)$ from
each data set are different.  To see this, we also plot contours of
constant $w_X$ which gives minimum values of $\chi^2$ at each point on
the $\Omega_m$ vs. $\Omega_X$ plane in
Fig.~\ref{fig:Om_Ox_wmarg_w0min}.  We can clearly see that the favored
values of $w_X$ vary for different data sets.  Observations we
consider here measure some distance scales to certain redshifts which
are determined by the energy density of matter $\Omega_m$, dark energy
$\Omega_X$ and the equation of state $w_X$.  If we vary the value of
$w_X$, the density parameters $\Omega_m$ and $\Omega_X$ can have more
freedom to be consistent with observations since the fit to the data
depends on the combinations of these quantities.  For a single
observation, when $w_X$ is allowed to vary, larger range of
$(\Omega_m, \Omega_X)$ can be consistent with observations.  However,
when we use all observations, the combinations of $\Omega_m$,
$\Omega_X$ and $w_X$ which are consistent with observations become
fairly limited.  Thus while the allowed regions become larger for
each data set, when we combine all the data, the allowed region
converges towards the concordance model with
$\Omega_m+\Omega_X\sim1$ and $\Omega_m \sim 0.3$ even if we do not
assume the cosmological constant as dark energy.

\begin{figure}[t]
    \centerline{\scalebox{1.2}{\includegraphics{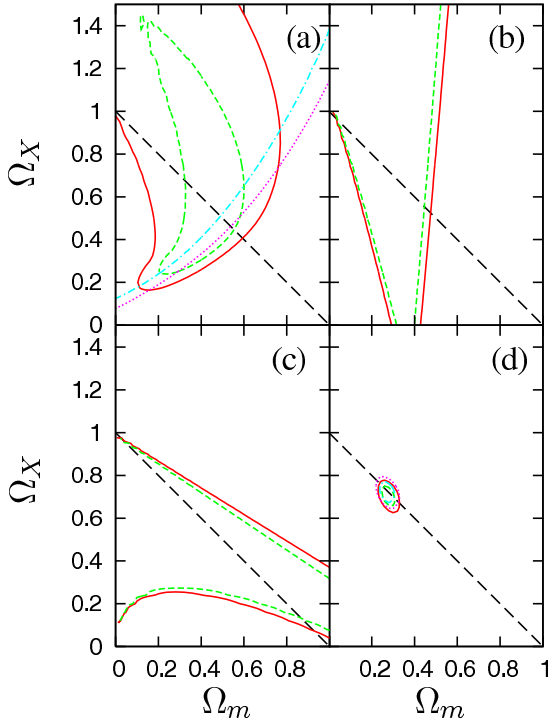}}}
	\vspace{0.5cm}
	\caption{The same as Fig.~\ref{fig:Om_Ox_wm1} except that we
	consider the time dependent equation of state for dark energy
	and marginalize over $w_0$ and $w_1$.}
\label{fig:Om_Ox_w01marg}
\end{figure}

Next we discuss the case with the time dependent equation of state
parameterized as Eq.~(\ref{eq:eos}).  As it has been already pointed
out in the literature, even if we assume a flat universe, there exists
a degeneracy among the parameters which describe the evolution of
equation of state for dark energy using a single observational
data. It is also known that the degeneracy can be removed using more
than one observation assuming the flatness.  Here we discuss to what
extent the evolution of dark energy equation of state can affect the
determination of $\Omega_m$ and $\Omega_X$ when we do not assume a
flat universe.  In Fig.~\ref{fig:Om_Ox_w01marg}, the contours of
1$\sigma$ and 2$\sigma$ constraints are shown as the same manner as
Fig.~\ref{fig:Om_Ox_wm1} except that we varied the both values of
$w_0$ and $w_1$ which appear in Eq.~(\ref{eq:eos}) and marginalized
over them to obtain the constraint. We assumed the prior on them as $
-5 \le w_0 \le 0$ and $-4 \le w_1 \le 4$ under the condition of
Eq.~(\ref{eq:w0w1}).  For the constraint from SNeIa, the allowed
regions get larger compared to the case with the constant equation of
state discussed above (see Fig.~\ref{fig:Om_Ox_wmarg}).  As for the
constraint from the baryon acoustic oscillation peak and the CMB shift
parameter, the allowed regions are almost the same as the case with a
constant equation of state.  For observations which measure a single
distance scale, the fit to the data does not significantly become
better, even if the time evolution of $w_X$ is allowed, due to the
degeneracy between $w_0$ and $w_1$.  Furthermore, when we use all
three different observations together, the allowed region becomes
almost the same as that of the case with a constant equation of state.
This is because each observation favors different values of $w_X$.
Even if we consider the time-varying equation of state, different
combinations of $w_0$ and $w_1$ are chosen to minimize the value of
$\chi^2$ for each observation. Thus the allowed region from the
combined data set is almost unchanged although the constraints from a
single observation, in particular from SNeIa, can become weaker.

Here we comment on the effect of the prior on $w_0$ and $w_1$.
Although the constraints from a single data alone somewhat depend on
the prior, the allowed region for all data combined is almost
unaffected since the combinations of $w_0$ and $w_1$ favored around
the allowed region from all data sets are far from the the edge of
the prior we assumed.

\section{Constraints on the $\Omega_m$ vs. $w_X$ plane}

\begin{figure}[t]
    \centerline{\scalebox{1.2}{\includegraphics{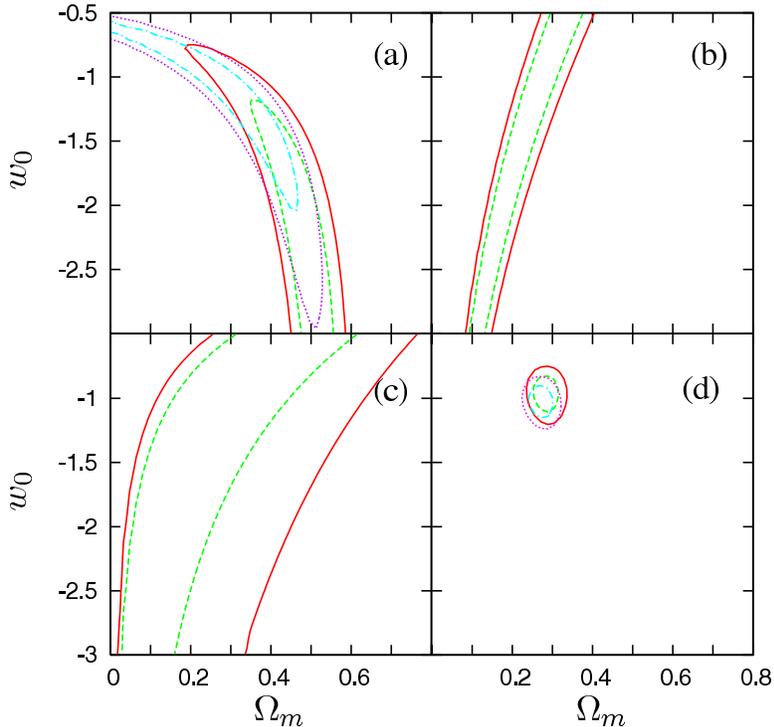}}}
	\caption{ Constraints on $\Omega_m$ and $w_X$ from
	observations of SNeIa (a), the baryon acoustic oscillation peak
	(b), CMB (c) and all data combined (d).  We assumed a flat
	universe and a constant equation of state here.}
\label{fig:Om_w0_flat}
\end{figure}

In this section, we consider the constraints on the $\Omega_m$
vs. $w_X$ plane.  First we show the constraint on $\Omega_m$ and $w_0$
in a flat universe with the equation of state for dark energy being
constant. In Fig.~\ref{fig:Om_w0_flat}, contours of 1$\sigma$ and
2$\sigma$ constraints from observations of SNeIa (a), the baryon
acoustic oscillation peak (b), the CMB shift parameter (c) and all
data combined (d) are shown.  As is well-known, SNeIa and CMB are
complementary for constraining dark energy, which can be seen
from the figure.  In addition, we can see that the constraint from the
baryon acoustic oscillation peak is also complementary.  Thus we can
obtain a severe constraint using all three data sets.

\begin{figure}[t]
    \centerline{\scalebox{1.2}{\includegraphics{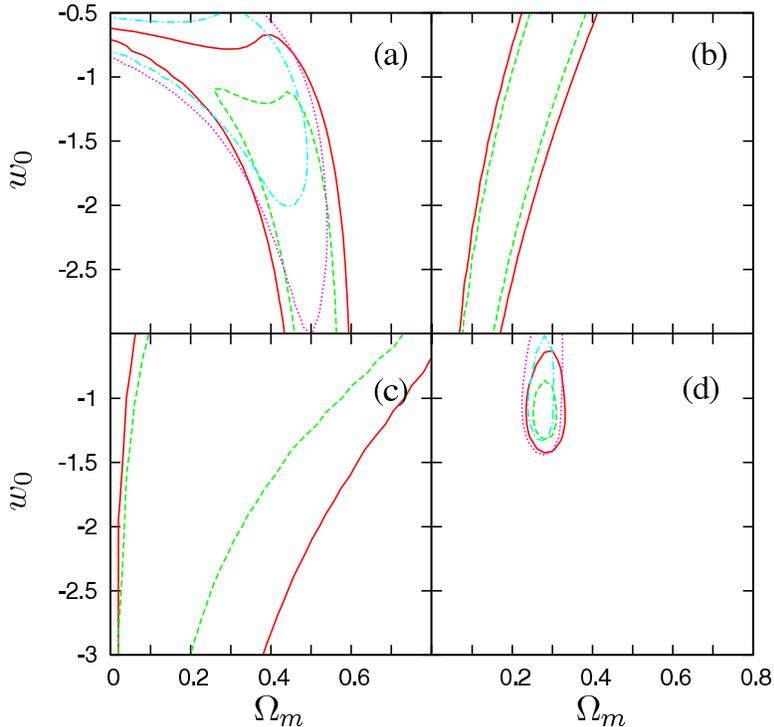}}}
	\caption{The same as Fig.~\ref{fig:Om_w0_flat} except that we
	consider the time dependent equation of state for dark energy
	as in Eq.~(\ref{eq:eos}).  To obtain the constraint, we
	marginalized over $w_1$.  Here we assumed a flat universe.}
\label{fig:Om_w0_flat_w1marg}
\end{figure}

Next we consider the case with the time-varying equation of state for
dark energy.  Here we discuss the case with a flat universe. In
Fig.~\ref{fig:Om_w0_flat_w1marg}, the constraints on $\Omega_m$ and
$w_0$ are shown marginalizing over $w_1$.  When we marginalize the
values of $w_1$, we assumed the prior on $w_1$ as $ -4 \le w_1 \le 4$.
Similarly to the situation where we constrain $\Omega_m$ and
$\Omega_X$ discussed in the previous section, if we consider the
time-varying equation of state for dark energy, the allowed region
becomes larger when we use a single data set. However, when we consider
all three data sets, we can obtain a relatively severe
constraint. Although the allowed region of $w_0$ extends to larger
values, $\Omega_m$ is constrained to be $\Omega_m\sim 0.3$ which is
the almost same as the case with assuming a constant equation of state.  This is
because the values of $w_1$ favored by different data sets given
$\Omega_m$ and $w_0$ are different. Thus the degeneracy among the
parameters describing the time dependence of the dark energy equation
of state can be removed drastically by using all three data sets.

\begin{figure}[t]
    \centerline{\scalebox{1.2}{\includegraphics{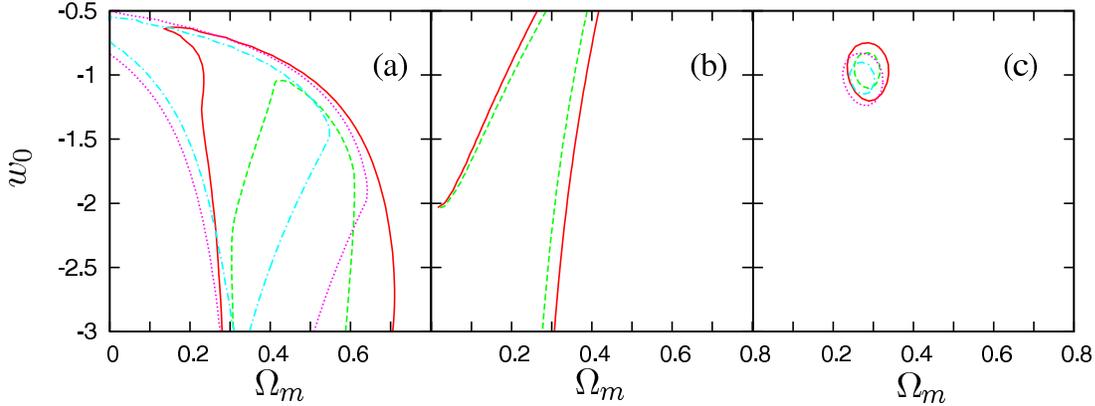}}}
	\caption{ Constraints on $\Omega_m$ and $w_X$ from
	observations of SNeIa (a), the baryon oscillation (b) and all data
	combined (c).  Here we do not assume a flat universe,  but a
	constant equation of state for dark energy is assumed.  In
	this figure, we marginalized over $\Omega_k$.}
\label{fig:Om_w0_Okmarg_w1fix}
\end{figure}

Now we discuss the constraints without assuming a flat universe.
First we discuss the case with $w_X$ being constant in time.  In
Fig.~\ref{fig:Om_w0_Okmarg_w1fix}, we show the constraints from SNeIa
(a), the baryon acoustic oscillation (b) and all data combined (c).
For the last case, the data from the CMB shift parameter is
included. To obtain the constraint, we marginalized over $\Omega_k$
with the prior $-0.3 \le \Omega_k \le 0.3$.  Here we do not show the
constraint from the CMB shift parameter alone since we cannot obtain a
significant constraint from it in the parameter region we consider in
Fig.~\ref{fig:Om_w0_Okmarg_w1fix}.  This is because, as it has already
been pointed out, the curvature of the universe and $w_X$ are strongly
degenerate in the CMB power spectrum \cite{Crooks:2003pa}. Again,
although the constraints from a single data set alone are weakened, if
we consider all data sets, we can obtain almost the same constraint as
that with the case where a flat universe is assumed. Notice that the
constraint on $w_0$ is also not changed much compared to that with
flat universe prior.  Hence from Fig.~\ref{fig:Om_w0_Okmarg_w1fix}, we
can say that the prior on the curvature does not affect the
determination of the equation of state for dark energy much.

\begin{figure}[t]
    \centerline{\scalebox{1.2}{\includegraphics{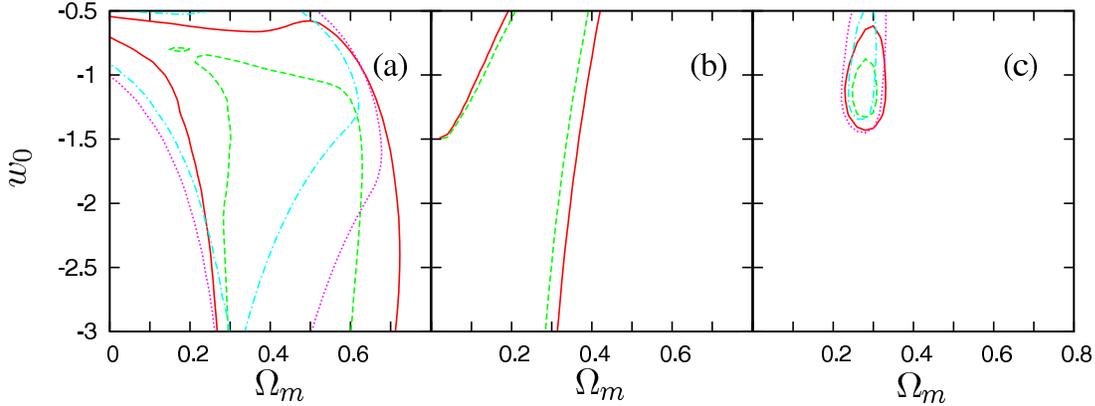}}}
	\caption{The same as Fig.~\ref{fig:Om_w0_Okmarg_w1fix} except
	that we consider the time dependent equation of state for dark
	energy and do not assume a flat universe.  In this figure, we
	marginalized over $\Omega_k$ and $w_1$.}
\label{fig:Om_w0_w1Okmarg}
\end{figure}

We also show the case where we consider the time-varying equation of
state without assuming a flat universe. In
Fig.~\ref{fig:Om_w0_w1Okmarg}, the constraints are shown as the same
as Fig.~\ref{fig:Om_w0_Okmarg_w1fix} except that we marginalized over
$\Omega_k$ and $w_1$ in this case.  The prior on $w_1$ is taken as $-4
\le w_1 \le 4$.  The CMB shift parameter cannot give a significant
constraint in the parameter range we consider in this case too.  As in
the previous cases, using each observational data alone, the allowed
region becomes significantly larger compared to those with constant
equation of state and a flat universe. However, when we use all data
sets, the allowed region does not change much. From
Figs.~\ref{fig:Om_w0_flat} and \ref{fig:Om_w0_Okmarg_w1fix}, we can
conclude that the prior on the curvature of the universe does not
affect the constraint on $\Omega_m$ and $w_0$ much if we use all data
sets.  In particular, we can obtain the constraint such that
$\Omega_m\sim 0.3$ without assuming a flat universe.  Moreover, the
assumption of the constancy of the equation of state for dark energy
also does not affect the constraint on $\Omega_m$. On the constraint
on the equation of state, if we consider the time-varying equation of
state, the allowed region becomes slightly larger even if all data
combined is used. 

\section{Constraints on the evolution of dark energy}

\begin{figure}[t]
    \centerline{\scalebox{1.2}{\includegraphics{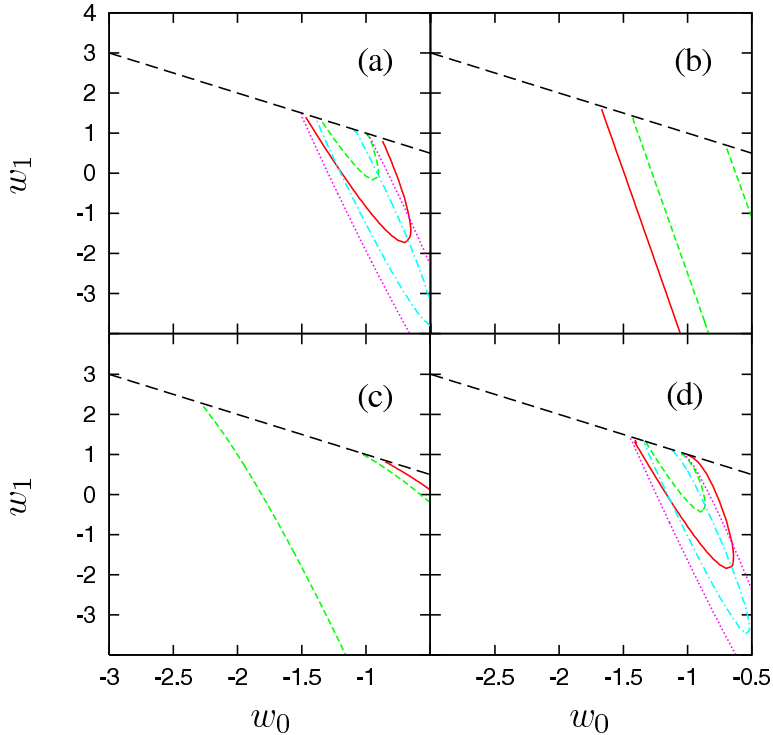}}}
	\caption{ Constraints on $w_0$ and $w_1$ from SNeIa (a),
	the baryon acoustic oscillation peak (b), the CMB shift parameter
	(c) and all data combined (d).  We assumed $\Omega_m=0.28$
	with a flat universe.}
\label{fig:w0_w1_flat}
\end{figure}

In this section, we consider the constraints on the time dependence of
dark energy equation of state, i.e., on $w_0$ and $w_1$ without
assuming the prior of a flat universe.  However first we discuss the
case with a flat universe for the later comparison with the case where
a flat universe is not assumed.  In Fig.~\ref{fig:w0_w1_flat},
contours of 1$\sigma$ and 2$\sigma$ allowed regions are shown for the
case with a flat universe. Here we fix the energy density of matter as
$\Omega_m=0.28$.  Notice that we do not consider the region where
$w_0+w_1 >0$ in order not to include the possibilities of early-time
dark energy domination.  As we can see from the figure, the constraint
from SNeIa is stringent compared to baryon acoustic oscillation peak
and CMB.  Since the position of the baryon acoustic oscillation peak
measures a single distance scale from $z=0$ to $z=0.35$ and the CMB
shift parameter also gives a single scale from $z=0$ to $z=1089$,
there is strong degeneracy among the parameters which describe the
equation of state for dark energy.  Notice that the distance scales
are determined by the integration of the inverse of the Hubble
parameter, which can smear out the information on the time dependence
of the equation of state. This is the reason why the strong degeneracy
exists when only a single scale is considered.  However, as for SNeIa
data, the degeneracy is removed to some extent since we have the
distances from $z=0$ to various redshifts\footnote{
Of course, there is a limitation to determine the time dependence of
the equation of state using SNeIa data
\cite{Maor:2000jy,Maor:2001ku}. }
. Thus observations of SNeIa can mainly constrain the time dependence of 
the equation of state for dark energy.

\begin{figure}[t]
    \centerline{\scalebox{1.2}{\includegraphics{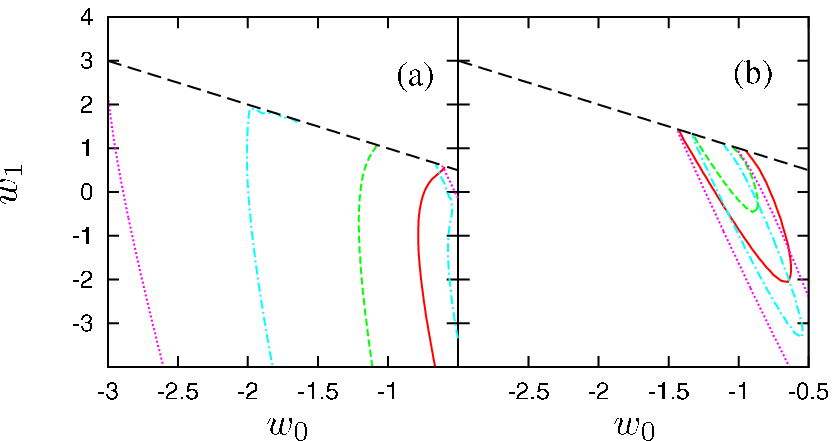}}}
	\caption{Constraints on $w_0$ and $w_1$ from observations of
	SNeIa (a) and all data combined (b). A flat universe is assumed but
	we marginalized over $\Omega_m$. }
\label{fig:w0_w1_flat_Ommarg}
\end{figure}

\begin{figure}[t]
    \centerline{\scalebox{1.2}{\includegraphics{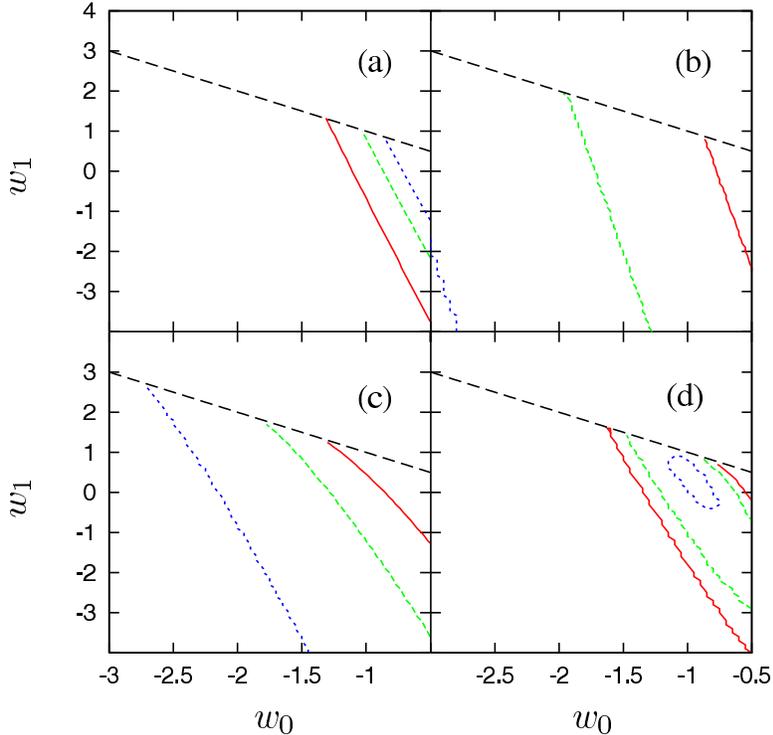}}}
	\caption{Contours of constant $\Omega_m$ which gives minimum
	$\chi^2$ when we marginalize over $\Omega_m$ for the
	constraints from SNeIa (a), the baryon acoustic oscillation peak
	(b), CMB shift parameter (c) and all data combined (d).
	Contours of $\Omega_m=0.3$ (red solid line), $0.2$ (green
	dashed line) and $0.1$ (blue dotted line) are shown except for
	the case where all three data are combined (panel (d)).  For
	the panel (d), $\Omega_m=0.29$ (red solid line), $0.28$ (green
	dashed line) and $0.27$ (blue dotted line) are shown.  For
	SNeIa, we use the data from SNLS.}
\label{fig:w0_w1_flat_Ommarg_minOm}
\end{figure}

Next we show the constraints on $w_0$ and $w_1$ in a flat universe
without assuming a particular value for $\Omega_m$.  To obtain the
constraint, we marginalized over $\Omega_m$ with the prior $0 \le
\Omega_m \le 0.5$. In Fig.~\ref{fig:w0_w1_flat_Ommarg}, we show the
constraints on $w_0$ and $w_1$ from SNeIa (a) and all data combined
(b).  We do not show the constraints from the baryon acoustic
oscillation peak and the CMB shift parameter because we cannot obtain
significant constraints from them in the parameter region of $w_0$ and
$w_1$ we consider here. Also in this case, such severe degeneracies
among the parameters exist because those observations measure a single
distance scale.  As in the previous case, when we include all data
sets in the analysis, we can obtain almost the same constraint as that
with $\Omega_m$ being fixed to $0.28$.  This is partly because the
best fit value of $\Omega_m$ is near $0.28$ when we marginalized over
it, but also because the values of $\Omega_m$ minimizing $\chi^2$ for
each observational data set are different.  To see this, in
Fig.~\ref{fig:w0_w1_flat_Ommarg_minOm}, we show contours of constant
$\Omega_m$ which minimize $\chi^2$ when we marginalize over $\Omega_m$
for the constraints from SNeIa (a), the baryon acoustic oscillation
peak (b), the CMB shift parameter (c) and all data combined (d).  As
is clear from the figure, preferred values of $\Omega_m$ for each
observation are quite different.  Thus when all data are used, we can
obtain a severe constraint although a single observation cannot
constrain the equation of state for dark energy. It should be also
mentioned that the prior on $\Omega_m$ does not affect the constraint
on the time dependence of the equation of state when we use all data
combined.

\begin{figure}[t]
    \centerline{\scalebox{1.2}{\includegraphics{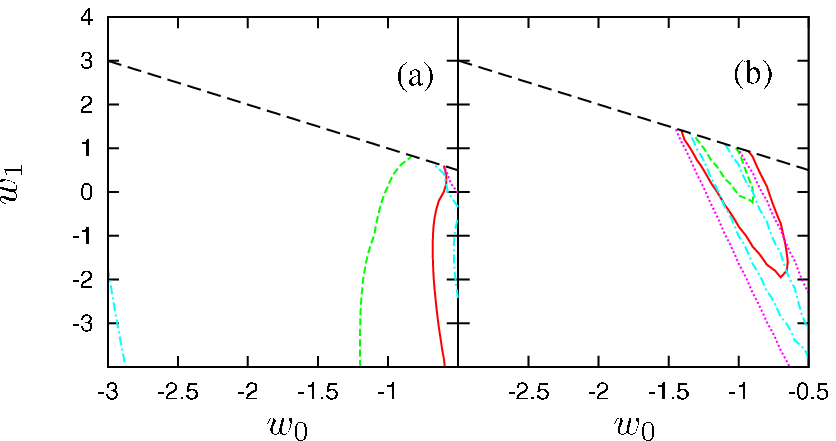}}}
	\caption{The same as Fig.~\ref{fig:w0_w1_flat_Ommarg} except
	that we marginalized over $\Omega_m$ and $\Omega_k$.  }
\label{fig:w0_w1_OmOkmarg}
\end{figure}

Now we discuss the case with $\Omega_m$ not being fixed and not
assuming a flat universe.  In Fig.~\ref{fig:w0_w1_OmOkmarg}, the
allowed regions are shown after marginalizing over $\Omega_m$ and
$\Omega_k$.  We assumed the prior on these variables as $ 0 \le
\Omega_m \le 0.5$ and $- 0.3 \le \Omega_k \le 0.3$.  In this case, the
constraint is significantly weakened when we consider a single data
set alone. We only report here the constraint from SNeIa (a) and that
from all data combined (b) since the baryon acoustic oscillation peak
and the CMB shift parameter cannot give meaningful constraints on this
plane in this case too.  As we can see, when we consider the
constraint from all data sets, it significantly becomes severe
compared to that from SNeIa alone as seen from the figure.  This is
because the favored values of $\Omega_m$ and $\Omega_k$ from each data
set are different as in the previous cases.  Thus we can conclude that
the prior on the curvature does not affect much the determination of
the equation of state for dark energy if we use all data combined.

\section{Summary}

We considered the constraint on the curvature of the universe and the
equation of state for dark energy from observations of SNeIa, the
baryon acoustic oscillation peak and the CMB shift parameter. Usually,
when one discusses the curvature of the universe, dark energy is
assumed to be the cosmological constant.  Moreover, when one considers
the constraint on the evolution of dark energy, in particular the time
dependence of the dark energy equation of state, a flat universe is
usually assumed.  In this paper, we discussed the constraints on the
curvature of the universe without assuming the cosmological constant
and also the time dependence of the equation of state for dark energy
without assuming a flat universe. We showed that the constraint on the
curvature of the universe is significantly relaxed from a single
observation when we allow the time dependence in the dark energy
equation of state. However, it was also shown that, when we use all
data sets, the curvature of the universe or the energy density of
matter and dark energy are severely constrained to be $\Omega_m \sim
0.3$ with a flat universe even if we consider the time-varying
equation of state for dark energy.  Observations we consider here
measure some distance scales to certain redshifts which can be
determined by the energy density of matter $\Omega_m$, dark energy
$\Omega_X$ and the equation of state $w_X$.  If we assume a broad
range for $w_X$, the energy density of matter $\Omega_m$ and dark
energy $\Omega_X$ can have more freedom to be consistent with
observations since the fit to the data depends on the combinations of
these variables.  For a single observation, if $w_X$ is allowed to
vary more freely, much larger range of values of $\Omega_m$ and
$\Omega_X$ can be consistent with observations.  However, when we use
all observations, the combinations of $\Omega_m$, $\Omega_X$ and $w_X$
consistent with observations become fairly limited.  Thus even if the
allowed regions become large for each data set, when we combine all
the data, we can obtain a severe constraint, which is interestingly
almost the same region as that we obtain assuming the cosmological
constant as dark energy.

We also investigated the constraint on the time-varying equation of
state for dark energy without assuming a flat universe. Similarly to
the situation where we constrained the curvature with the time-varying
equation of state, the allowed region for $w_X$ becomes larger when we
use a single observational data.  However, if we use all data sets
considered in this paper, we can obtain almost the same constraint as
that in the case where a flat universe is assumed.

Finally, we summarize what we found in this paper. The combination of the current 
observations 
\begin{itemize}

\item  favors a flat universe regardless of the prior on the equation of state
for dark energy with or without time evolution.

\item favors $\Omega_m \sim 0.3$  regardless of the flatness prior and 
the prior on the  equation of state
for dark energy with or without time evolution.

\item yields constraints on the time evolution of dark energy equation 
of state regardless of the prior on $\Omega_m$ and $\Omega_k$.  

\end{itemize}

In future observations, we can obtain much more stringent
constraints on the equation of state for dark energy as well as the
curvature of the universe regardless of the prior on other
cosmological parameters. Hence we can expect that we will be able to
have much insight on the nature of dark energy and also the
inflationary paradigm.

\bigskip
\bigskip

\noindent
{\bf Acknowledgment:} T.T would like to thank the Japan Society for
Promotion of Science for financial support.

\end{document}